\documentclass{rnaastex}

\begin{document}

\title{Detection of an OH 1665-MHz Maser in M33}

\correspondingauthor{Eric Koch}
\email{ekoch@ualberta.ca}
\author[0000-0001-9605-780X]{Eric Koch}
\author[0000-0002-5204-2259]{Erik Rosolowsky}
\affiliation{University of Alberta, Department of Physics 4-183 CCIS, Edmonton AB T6G 2E1, Canada}

\author{Megan C.\ Johnson}
\affiliation{United States Naval Observatory, 3450 Massachusetts Ave NW, Washington, D.C., 20392, USA}

\author{Amanda A.\ Kepley}
\affiliation{National Radio Astronomy Observatory, 520 Edgemont Road, Charlottesville, VA 22903-2475, USA}

\author{Adam Leroy}
\affiliation{The Ohio State University, Department of Astronomy, 140 West 18th Avenue, Columbus, OH 43210, USA}

\keywords{processes: masers --- galaxies: individual (M33)}


\section{}

Galactic hydroxyl (OH) masers arise from regions surrounding late-type stars and young stellar objects \citep{gray2012}.  These masers provide an important measure of the environment of these regions.  Polarization measurements provide a direct measure of the magnetic field strengths, and their high surface brightness allows for detailed kinematics and distance determinations through Very-Long-Baseline Interferometry \citep{fish2006}.  Different observed ratios between main-line transitions (1665 \& 1667 MHz) and the satellite lines (1612 \& 1720 MHz) also provide clues to the origin of the emission, as different sources emit stronger in certain lines \citep{gray2012}.

While thousands of OH masers have been detected within the Galaxy \citep{engels2015} and OH megamasers have been found in distant galaxies \citep{darling2002}, far fewer have been detected in the Local Group.  In the LMC, several 1665-MHz masers were reported by \citet{brooks1997}, and \citet{marshall2004} find ten 1612-MHz maser sources.  A search for OH masers in M31 by \citet{willett2011} yielded no detections with a $5\sigma$ limit of $10\mbox{ mJy}$ in $4.4 \mbox{ km s}^{-1}$ channels. \citet{fix1985} performed a survey of M33 for OH masers and also found no detections at a $5\sigma=15\mbox{ mJy}$ limit within $5.2 \mbox{ km s}^{-1}$ channels.

We report one 1665-MHz detection in new observations with the NSF's Karl G. Jansky Very Large Array. To the authors' knowledge, this is the first OH maser detection in M33.  A full description of the data is given in Koch et al. (in prep).  We image all four OH lines in natural and uniform weighting with  $1.5\mbox{ km s}^{-1}$ channels and search by-eye for bright emission.  There are no detections in the other three lines.  We deconvolve the 1665-MHz data within a mask twice the uniform-weighted beam size ($9\farcs 7$) centered on the source with a threshold of $\sim2\sigma=3$ mJy~beam$^{-1}$.  The source is unresolved at this scale, corresponding to $\sim40$ pc at the distance of M33 \citep[840 kpc;][]{freedman2001}.

\begin{figure}[htp!]
\begin{center}
\includegraphics[scale=0.65,angle=0]{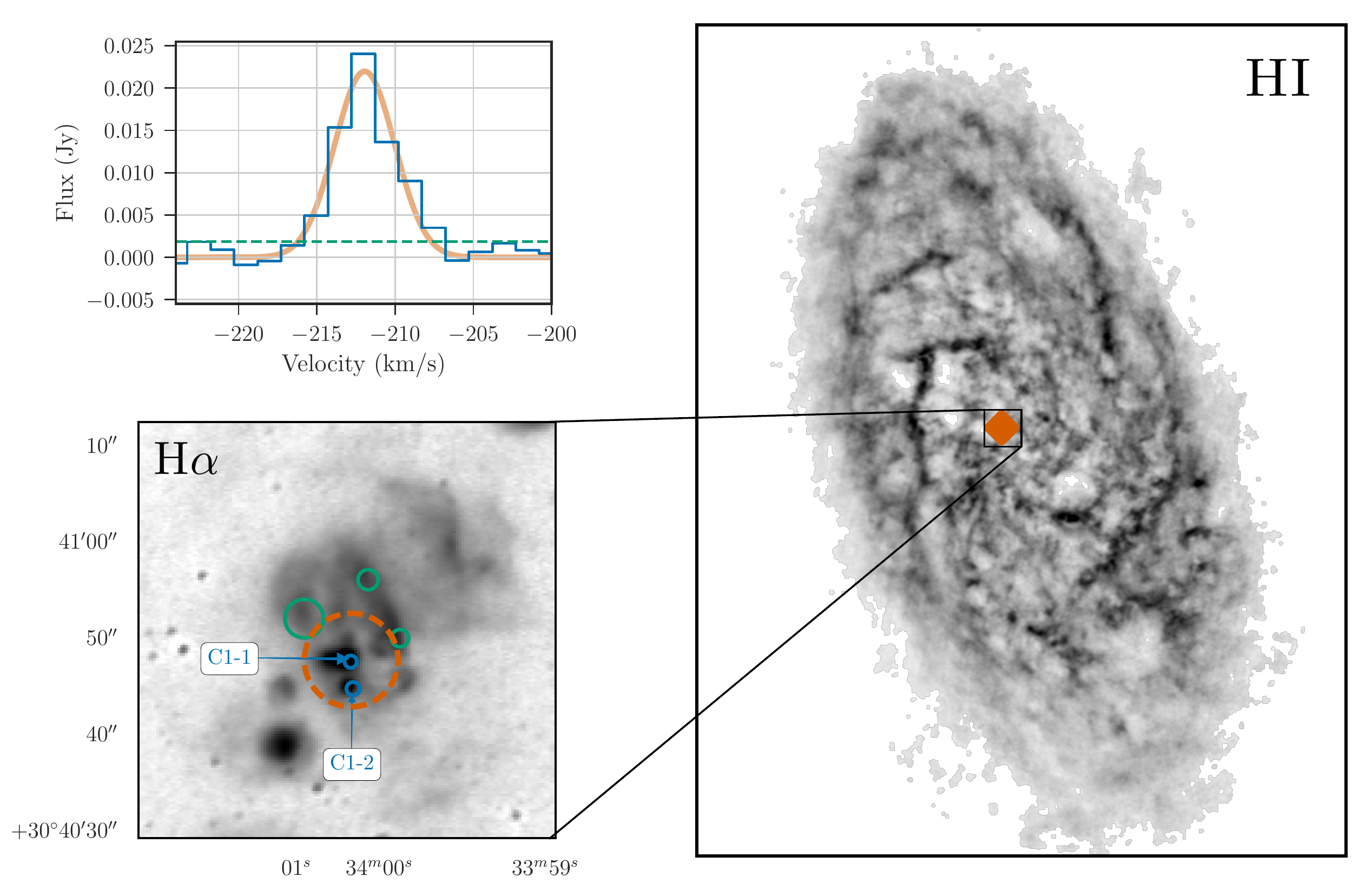}
\caption{\label{fig:1} Right: Position of the OH detection on an integrated 21-cm map of M33 (Koch et al., in prep). Top Left: Spectrum at the peak intensity position. The dashed green line is the noise level at 1.85 mJy and the thick orange line is a Gaussian fit. Bottom Left: H$\alpha$ image of the region shown in the right panel \citep{hodge1999}.  Sources from the \citet{moody2017} catalogue are shown in blue within the beam area, and green for nearby regions [from left to right, C1-10 \citep{moody2017}, 79a and 79c \citep{boul1974}].}
\end{center}
\end{figure}

Figure \ref{fig:1} shows the spectrum of the source, which is adequately modeled as a Gaussian with a FWHM of $4.5\pm1.8\mbox{ km s}^{-1}$. However, the model has significant residuals ($2\sigma$) at the peak, suggesting the true spectrum is more complex.  A spectrum comprised of multiple (unresolved) narrow OH features is consistent with the typical $<1\mbox{ km s}^{-1}$ FWHMs measured for Galactic OH masers \citep{argon2000,fish2006}.  The peak flux density is 24.1 mJy at $V_{\mathrm{LSRK}}=-212.0\pm0.8\mbox{ km s}^{-1}$.  The integrated flux over the line profile is $110.6 \mbox{ Jy m s}^{-1}$. This corresponds to a luminosity of $5.2\times10^{29} \mbox{ erg s}^{-1}$ at the distance of M33.  By down-sampling our data to $5.2 \mbox{ km s}^{-1}$, we find this source would correspond to a $2.8\sigma$ detection in the \citet{fix1985} data.

We fit a two-dimensional Gaussian profile to the integrated intensity map over a velocity range of $-206$ to $-218 \mbox{ km s}^{-1}$.  The fit constrains the peak location to $\alpha=23.50069(9)^{\circ}$ and $\delta = 30.67984(9)^{\circ}$ (J2000), as shown in Figure \ref{fig:1}.  Using the catalog from \citet{moody2017}, two H$\alpha$ sources fall within the beam area: C1-1 and C1-2. The former source is labeled as 79b in \citet{boul1974}.  \citet{johnson2001} classify this region as a potential ultra-compact \ion{H}{2} region; the source is the second brightest radio source in their sample for M33.  \citet{moody2017} construct an SED for C1-1 from the UV to IR that instead suggests the source is the variable star D33 J013400.3+304048.1. \citet{moody2017} find that the second source, C1-2, is consistent with the radio source Dixon IK 50.


The $5\sigma$ peak flux density limit in our data is $1.85\mbox{ mJy}$ in a 1.5 km~s$^{-1}$ channel.  This corresponds to a specific luminosity of $1.65\times10^4\mbox{ Jy kpc}^{2}$.  The fact that this maser is the only detection, despite its high significance, suggests it is bright compared to Galactic OH masers. This is consistent with the upper limit of the Galactic OH 1665-MHz luminosity function from \citet{caswell1987}, where $1.65\times10^4 \,\mbox{Jy kpc}^{2}$ corresponds to the $99$\textsuperscript{th} percentile.



\acknowledgments

EWK is supported by a Postgraduate Scholarship from the Natural Sciences and Engineering Research Council of Canada (NSERC). EWR acknowledges the support of NSERC, funding reference number RGPIN-2017-03987. The National Radio Astronomy Observatory is a facility of the National Science Foundation operated under cooperative agreement by Associated Universities, Inc.

\software{CASA 5.1, radio-astro-tools (\url{http://radio-astro-tools.github.io}), aplpy \citep{robitaille}}

\end{document}